\documentclass[runningheads]{llncs}

\usepackage{tikz}
\usepackage{verbatim}
\usepackage{pgfplots}
\pgfplotsset{compat=1.12}
\usepackage{color, colortbl}

\usepackage[utf8]{inputenc} 
\usepackage[T1]{fontenc}    
\usepackage{url}            
\usepackage{booktabs}       
\usepackage{nicefrac}       
\usepackage{microtype}      
\usepackage{units}
\usepackage{dsfont}
\usepackage{graphicx}
\usepackage{booktabs}
\usepackage{commath}
\usepackage{enumitem}
\usepackage{xargs}
\usepackage{blindtext}
\usepackage{tabularx}
\usepackage{multirow}
\usepackage{multicol}
\usepackage{verbatim}
\usepackage{arydshln}

\usepackage{nameref}
\usepackage{varioref}

\usepackage{hyperref}       
\hypersetup{
	colorlinks = true,
	linkcolor = blue,
	anchorcolor = blue,
	citecolor = blue,
	filecolor = blue,
	urlcolor = blue
}
\usepackage{cleveref}

\graphicspath{ {img/} }

\newcommand{\ie}{{\it i.e.}\@\xspace}
\newcommand{\eg}{{\it e.g.}\@\xspace}

\usetikzlibrary{arrows,shapes,snakes,automata,backgrounds,petri,calc}
\usetikzlibrary{shapes,backgrounds, arrows.meta}

\usepackage[colorinlistoftodos,prependcaption,textsize=smal]{todonotes}
\newcommandx{\unsure}[2][1=]{\todo[linecolor=red,backgroundcolor=red!25,bordercolor=red,#1]{#2}}
\newcommandx{\change}[2][1=]{\todo[linecolor=blue,backgroundcolor=blue!25,bordercolor=blue,#1]{#2}}
\newcommandx{\info}[2][1=]{\todo[linecolor=OliveGreen,backgroundcolor=OliveGreen!25,bordercolor=OliveGreen,#1]{#2}}
\newcommandx{\improvement}[2][1=]{\todo[linecolor=Plum,backgroundcolor=Plum!25,bordercolor=Plum,#1]{#2}}
\newcommandx{\thiswillnotshow}[2][1=]{\todo[disable,#1]{#2}}

\usepackage{algorithm}
\usepackage{algpseudocode}

\algblock{ParFor}{EndParFor}
\algnewcommand\algorithmicparfor{\textbf{parfor}}
\algnewcommand\algorithmicpardo{\textbf{do}}
\algnewcommand\algorithmicendparfor{\textbf{end\ parfor}}
\algrenewtext{ParFor}[1]{\algorithmicparfor\ #1\ \algorithmicpardo}
\algrenewtext{EndParFor}{\algorithmicendparfor}

\begin{document}
\title{Unsupervised Hierarchical Grouping of Knowledge Graph Entities}
%
%
\author{Sameh K. Mohamed\inst{1,2,3}} 
\authorrunning{S.K. Mohamed}
%
\institute{Data Science Institute \and 
Insight Centre for Data Analytics \and
National University of Ireland Galway\\
\email{sameh.mohamed@insight-centre.org}}
\maketitle              
\begin{abstract}
Knowledge graphs have attracted lots of attention in academic and industrial environments. Despite their usefulness, popular knowledge graphs suffer from incompleteness of information especially in their type assertions. This has encouraged research in the automatic discovery of entity types. In this context, multiple works were developed to utilise logical inference on ontologies and statistical machine learning methods to learn type assertion in knowledge graphs. However, these approaches suffer from limited performance on noisy data, limited scalability and the dependence on labelled training samples. In this work, we propose a new unsupervised approach that learns to categorise entities into a hierarchy of named groups. We show that our approach is able to effectively learn entity groups using a scalable procedure in noisy and sparse datasets. We experiment our approach on a set of popular knowledge graph benchmarking datasets, and we publish a collection of the outcome group hierarchies\footnote{\url{https://samehkamaleldin.github.io/kg-hierarchies-gallery/}}.

\end{abstract}

\section{Introduction}
\label{sec:introduction}

Type information of knowledge graph (KG) entities is an important feature that categorizes entities of similar semantics. It is used in different tasks related to knowledge graphs including meta path extraction~\cite{DBLP:conf/www/MengCMSZ15}, link prediction~\cite{DBLP:conf/semweb/KrompassBT15} and fact checking~\cite{DBLP:journals/kbs/ShiW16}. Naturally, knowledge entities are categorized into a hierarchically structured set of classes \eg, \textit{person $\Rightarrow$ artist $\Rightarrow$ singer}, or \textit{location $\Rightarrow$ country $\Rightarrow$ city}, where each class encloses entities of similar properties. Hierarchical class structure provides richer semantics of entity type information that can be used as a feature in different knowledge graph tasks \eg link prediction and entity linking~\cite{DBLP:conf/ijcai/XieLS16}. Despite their usefulness, famous knowledge graphs from different domains suffer from type assertion incompleteness~\cite{DBLP:conf/wims/MeloPV16}. For example, type assertions of DBpedia 3.8 is estimated to have at most an upper bound of completeness 63.7\%~\cite{DBLP:journals/ijswis/PaulheimB14}. YAGO2 types are also estimated to be at most 53.3\% complete~\cite{DBLP:journals/ijswis/PaulheimB14}. This incompleteness has motivated research into automatic discovery of knowledge graph entity types.

The currently available approaches for classifying knowledge graph entities can be classified into two categories. First, schema-based approaches, where the developed approaches utilise the known schema of the knowledge graph to learn entity types. This includes automatically inferring type information using standard RDF reasoning~\cite{DBLP:conf/rweb/PolleresHDU13}, which applies rules of logical inference to infer new type assertion statements from existing ones using knowledge graph schema information. However, this approach is sensitive to noisy information, and it depends on prior-defined ontologies, and its predictions are bounded by the set of types defined by the ontology. The SDType~\cite{DBLP:conf/semweb/PaulheimB13} model is another schema-based approach which introduced using the link based type inference approach which depends on the assumption that relations happen between particular types. For example, if  entities $e_1$ and  $e_2$ are connected with relation "\textit{LiveIn}", then we can infer that $e_1$ is a person and $e_2$ is a place using relation defined domain and range types from schema. This approach provided efficient classification of entities in noisy knowledge graphs. However, it depended on the presence of schema information, and it also did not provide hierarchical structure of type information.

Secondly, the statistical learning approaches, which treat type prediction task as a multi-label classification problem, where models learn entity types using a set of graph based features. They use known type assertions as training examples and learn a model that can infer other unknown type assertions. They are known to provide more robust-to-noise type predictions~\cite{DBLP:journals/ijswis/PaulheimB14} compared to schema-based and logic-based approaches. These models  can also infer type hierarchies by using hierarchical multi-label classifiers~\cite{DBLP:conf/wims/MeloPV16}. However, these models require training assertions and it can not suggest new entity types.

\begin{figure}[t]
	\centering
	\resizebox{0.88\textwidth}{!}{
		
		\includegraphics{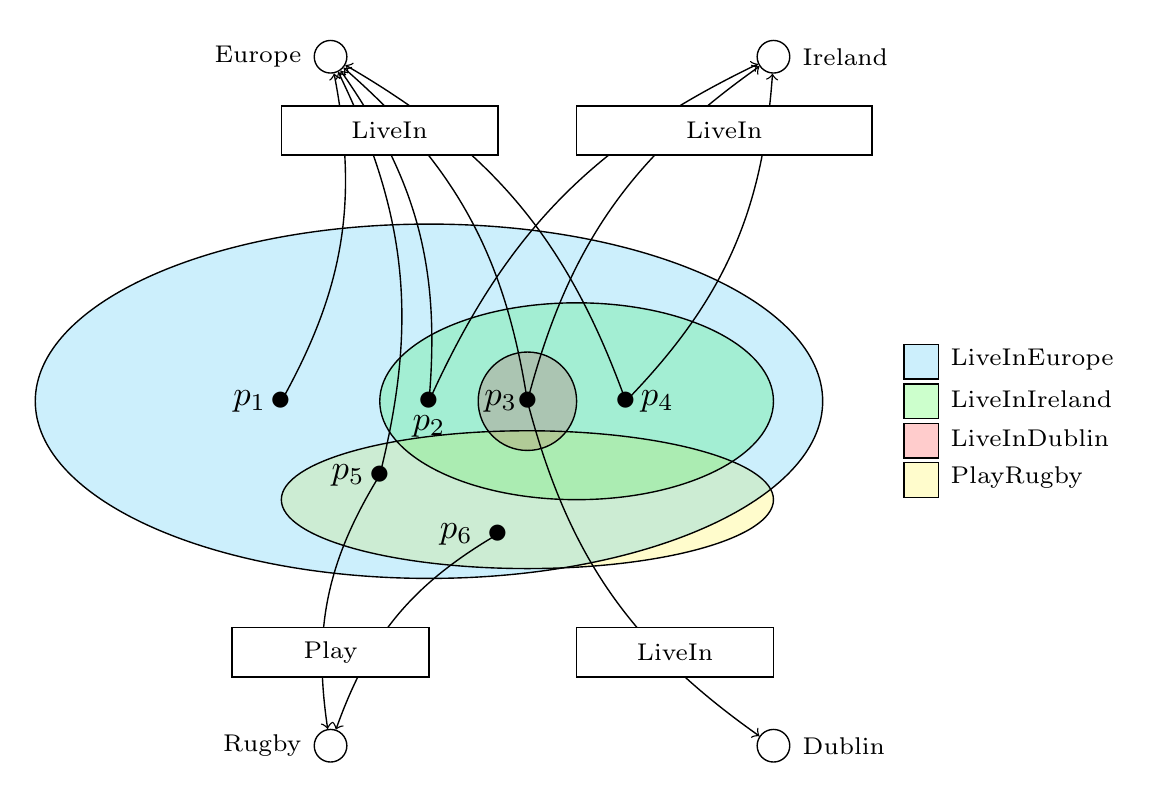}
	}
	\caption{Venn diagram of groups of people from a sample of facts about people living in Europe}
	\label{fig:venn-diagram}
\end{figure}

In this work, we propose a new unsupervised approach that groups knowledge graph entities according to their connected neighbour in the graph. This can be formulated such that for every set of entity nodes $E$ connected to another object node $e_o$ with a relation $r$, the set of nodes $E$ belongs to the group $g_{r,e_o}$. For example, in a general knowledge graph about people and cities all the people entities connected to the entity "\textit{Dublin}" with the relation "\textit{live-in}" are associated with the group "\textit{live-in-Dublin}" as shown in Fig.~\ref{fig:venn-diagram}. We then learn the hierarchy of these groups using intersection and containment ratios between them to build a hierarchy of entity types. This allows us to produce new named groups for entities, and provides rich hierarchy of the newly created groups.

The rest of this work is organised as follows:
\begin{enumerate}[label=(\Alph*)]
	\item We present a detailed description of the pipeline of our approach in Section~\ref{sec:model}.
	\item We experiment our method on multiple popular knowledge graph benchmarking datasets and we show samples of produced hierarchies in Section~\ref{sec:experiments}.
	\item We discuss related works in Section~\ref{sec:related}.
	\item We present our conclusions and future works in Section~\ref{sec:conc}.
\end{enumerate}
\section{Methods}
\label{sec:model}

In this section, we present our approach for learning hierarchical groups of entities in knowledge graphs. We divide the pipeline of our method into three segments: (1) entity grouping, (2) computing group similarities and (3) building group hierarchy. In the following subsection, we discuss the motivation for our approach and we discuss also each of its pipeline segments in detail while we use the knowledge graph sample in Fig.~\ref{fig:venn-diagram} as a running example.

\subsection{Motivation}
Knowledge graphs can model different types of assertions depending on their predicate types such as attribute assertions like the age or names of an entity, or related entities like birthplaces, friend, etc. Usually, the cardinality of relational predicates are many-to-many like having a friend, associated nationality or working for an organisation. In this case, our brains intuitively group entities associated with same predicates and destinations to a group like "friend of Jack", "people with British nationality " or "companies working in IT". In our approach, we use the same technique, where we transform every knowledge assertion \ie SPO triple to a group assertion ("S" belongs to group "PO"). We also use a configurable minimum size requirement for groups to avoid including one-to-one predicates as group assertions. 

\subsection{Generating Entity Groups}
Given any knowledge graph, our approach starts with generating groups of entities by transforming the graph facts into group assertions as previously discussed. For example, the knowledge graph in Fig.~\ref{fig:venn-diagram} contains multiple facts about six persons. In this graph, all the persons are living in Europe such that $\forall p \in \{p_1, ..., p_6\} (p, "LiveIn", "Europe")$. This is transformed to creating a group called "\textit{LiveIn\_Europe}" ($g$) such that $g=\{p_1, p_2,p_3,p_4,p_5,p_6\}$. Similarly, other groups are created like "\textit{LiveIn\_Ireland}"$\mapsto\{p_2,p_3,p_4\}$, "\textit{LiveIn\_Dublin}" $\mapsto\{p_3\}$ and "\textit{Play\_Rugby}"$\mapsto\{p_5,p_6\}$. This procedure is performed to generate groups from all the triples in the knowledge graph. 

Algorithm~\ref{alg:gen_groups} describes the process of generating these groups, where processing facts is parallelised to speed up processing large volumes of data. First, the set of all knowledge graph triples is divided into a configurable $j$ number of splits. Each of these splits is then processed to generate a dictionary of groups and their contained entities according to their own set of triples. All the resulting dictionaries are then joined to generate one dictionary of all groups in the graph and their corresponding entities. In order to restrict the extracted groups to a minimum specific number of entities, groups with size less than a configurable minimum size are removed from the group dictionary. Finally, the outcome of this procedure is a dictionary of the remaining groups and their member entities.

\begin{algorithm}[t]
	\caption{Generating entity groups in a knowledge graph}
	\label{alg:gen_groups}
	\begin{algorithmic}[1]
		\Require group min. size $\alpha$, KG triplets $T$, num. of jobs $j$, group dictionary $\mathcal{D}$
		\State initialise ${d_1, d2, ..., d_j}$ as empty dictionaries
		\State $t_1,t_2,...,t_j \gets \mathrm{split}(T, j)$
		\ParFor{$i \gets {[1, 2, ..., j]} $}
		\For{$(s, p, o) \in t_i$}
		\State $\mathrm{group} \gets \mathrm{concatenate}(p,o)$
		\If{group in $d_i$}
		\State $d_i[\mathrm{group}].\mathrm{append}(s)$
		\Else
		\State $d_i[\mathrm{group}] \gets [s]$
		\EndIf
		\EndFor
		\EndParFor
		\For{$i \gets {[1, 2, ..., j]} $}
		\State $\mathcal{D} \gets \mathcal{D} + d_i$
		\EndFor
		\For{$g \in \mathcal{D}$}
		\If{$\mathrm{size}(g) < \alpha$}
		delete $\mathcal{D}[g]$
		\EndIf
		\EndFor
		\State\Return{$D$}
	\end{algorithmic}
\end{algorithm}

\subsection{Computing Group Similarity}
We compute similarity measures between entity groups to learn their similarities and hierarchical structure. We compute two types of similarities to achieve that: Jaccard and hub promoted index (HPI) similarities~\cite{Lu2010LinkPI} which can be defined as follows:
\[
S^{jaccard}_{g1,g2} = \frac{\Gamma(g1)\cap\Gamma(g2)}{\Gamma(g1)\cup\Gamma(g2)} \quad,\quad S^{HPI}_{g1,g2} = \frac{|\Gamma(g1)\cap\Gamma(g2)|}{\min(|g1|,|g2|)},
\]
for any two groups $g1$ and $g2$, where $\Gamma(g)$ denotes the set of member entities of the group $g$ and $|g|$ denotes its size. The HPI similarity in this context computes the overlap between two groups, where its maximum value $1$ implies that one of the groups is a subset of the other. The Jaccard similarity on the other hand computes the overall similarity between two groups of entities. For example, the HPI similarity between "\textit{LiveInEurope}" and "\textit{LiveInIreland}" is equal to ${|\{p2,p3,p4\}|}/{min(3,6)} = 1$, which implies that the small group "\textit{LiveInIreland}" is a subset of the larger group "\textit{LiveInEurope}".

We also compute the similarities in a parallel procedure similar to the generation of groups. We first generate all possible combinations of groups, then we divide these combinations into splits. We then compute similarities for each of the splits. the outcome similarities of all the parallel procedures are then joined to generate a similarity matrix between all the groups.

\subsection{Building Groups Hierarchy}
The range of the hub promoted index (HPI) similarity between two groups is bounded between $0$ and $1$, where $0$ represents that the groups are independent and 1 implies that one group is a subset of the other. In noisy knowledge graph, the HPI index between totally dependant groups is always less than $1$ due to missing members in one of the two groups. In our approach we use a configurable parameter $\theta$ that represents the group containment HPI threshold. We initialise this parameter with a value of $0.9$ by default to tolerate 10\% of information loss.

\section{Experiments}
\label{sec:experiments}
In this section, we describe the datasets and outcomes of the experimentation of our approach.

\subsection{Data}
In our experiments we use six knowledge graph benchmarking datasets:
\begin{itemize}
	\item { WN18 \& WN18RR}: subsets of the WordNet dataset~\cite{Miller:1995} which contain lexical information of the English language~\cite{DBLP:conf/nips/BordesUGWY13,DBLP:conf/aaai/DettmersMS018}.
	\item {FB13k}: a subset of the freebase dataset~\cite{DBLP:conf/sigmod/BollackerEPST08} which contains information about general human knowledge~\cite{DBLP:conf/nips/BordesUGWY13,DBLP:conf/emnlp/ToutanovaCPPCG15}.
	\item { YAGO10}: a subset of of YAGO3 dataset~\cite{DBLP:conf/cidr/MahdisoltaniBS15} which contains information mostly about people and their citizenship, gender, and profession knowledge~\cite{DBLP:conf/aaai/Bouchard15}.
	\item {NELL239}: a subset of NELL dataset~\cite{DBLP:journals/cacm/MitchellCHTYBCM18,DBLP:conf/emnlp/GardnerM15} which  contains general knowledge about people, places, sports teams, universities, etc.
\end{itemize}
The above mentioned datasets are divided into three splits: training, validation and testing. In our experiments, we first join the three splits and we execute our method on the full dataset.


\begin{figure}[t]
	\centering
	\resizebox{1.1\textwidth}{!}{
		
		\includegraphics[angle=270]{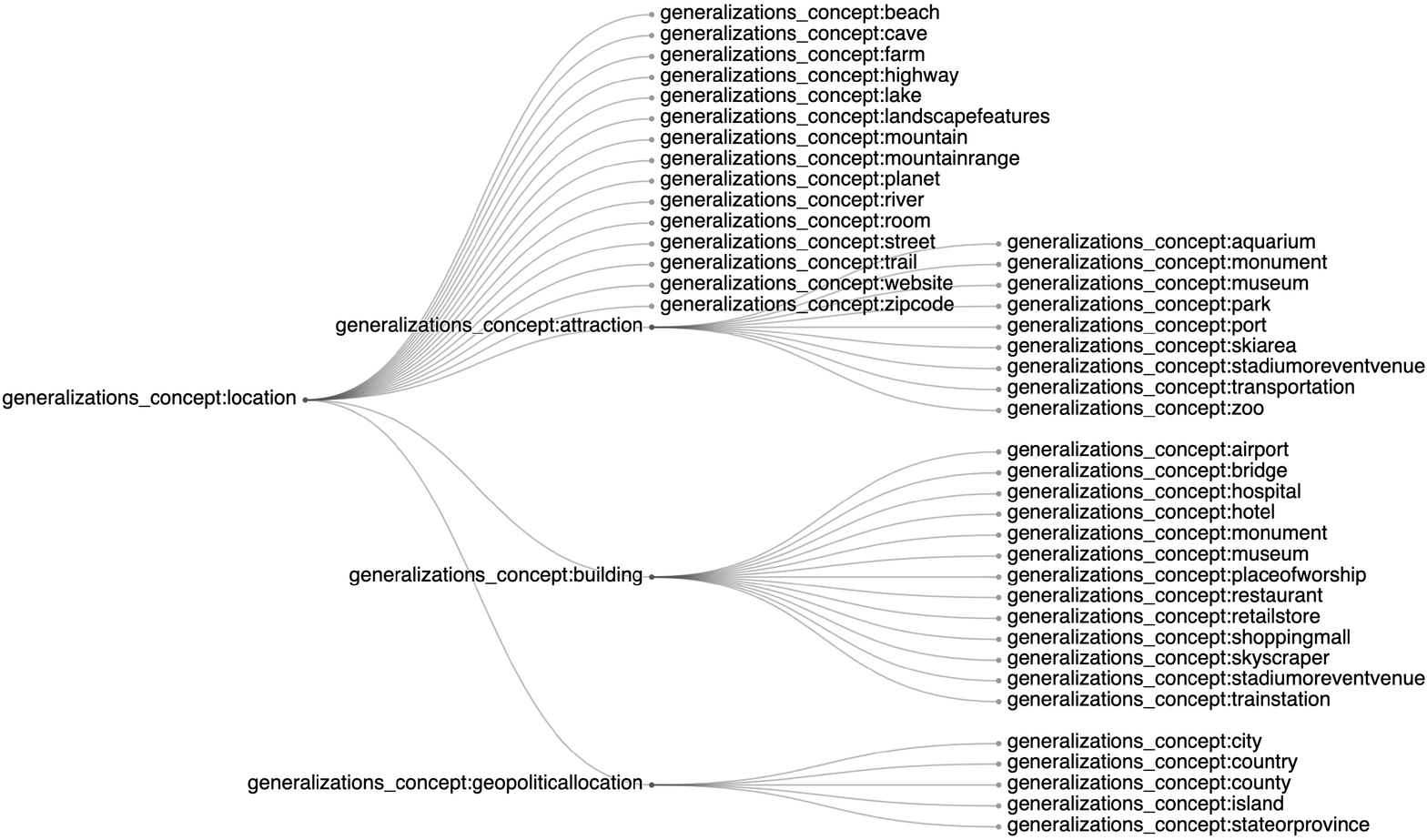}
	}
	\caption{An example of extracted location-based group hierarchy of a set of entities from the NELL239 dataset.}
	\label{fig:result_example_h}
\end{figure}
\begin{figure}[t]
	\centering
	\resizebox{0.99999\textwidth}{!}{
		
		\includegraphics[angle=270]{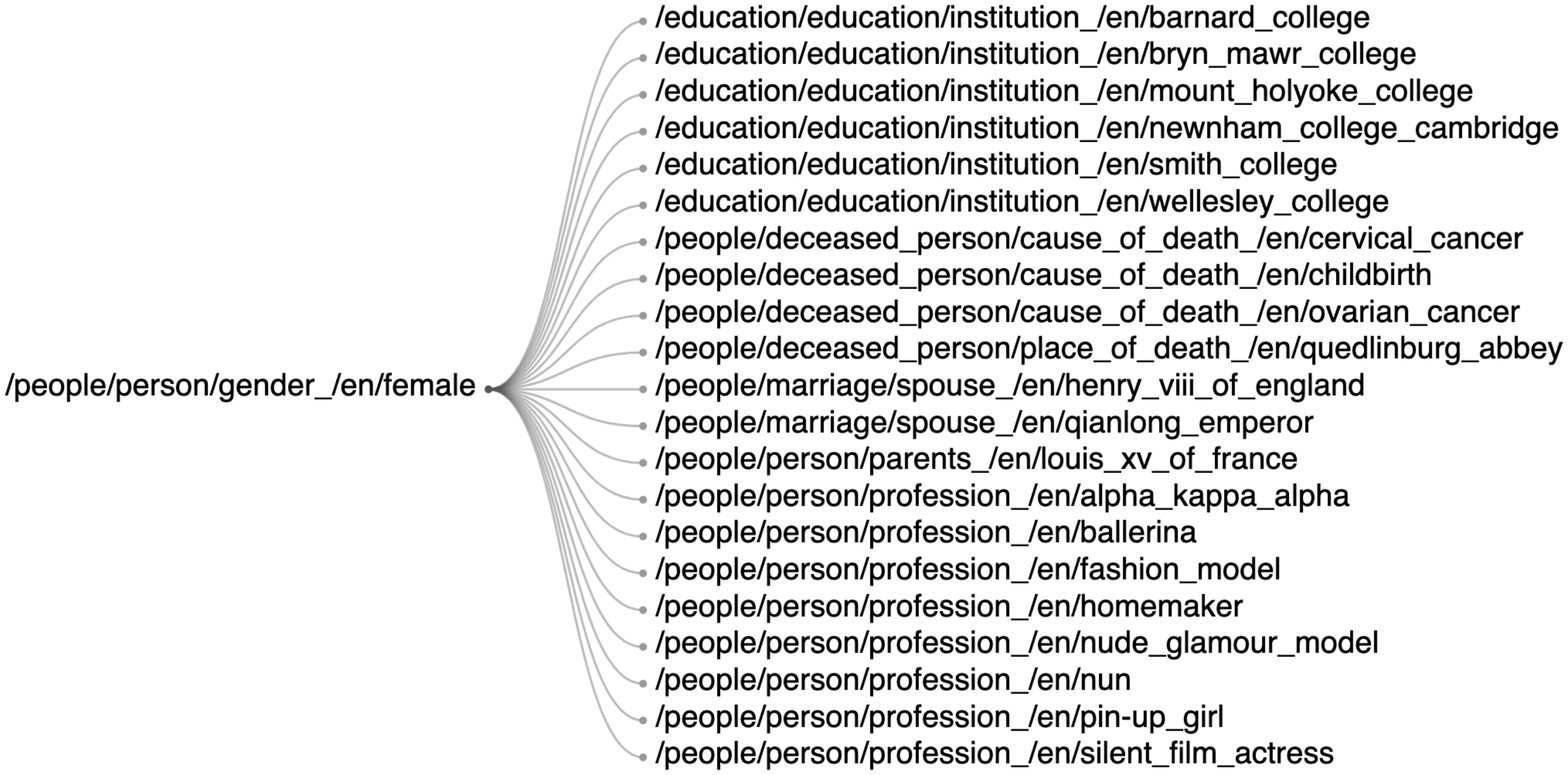}
	}
	\caption{An example of extracted gender-based group hierarchy of a set of entities for females from the FB13 dataset.}
	\label{fig:result_females}
\end{figure}
\subsection{Outcomes}
We executed our approaches on the aforementioned datasets, and we have generated the entity group hierarchies for each one of them. We use a minimum group size of $10$ and a threshold of $0.9$ for the HPI similarity for all the experimented datasets.

The outcome results show that these datasets contain different root groups, where each root include super group of different group semantics such locations, people and organisations.

Fig.~\ref{fig:result_example_h} shows an example of an outcome hierarchy of location-based entities in the NELL239 dataset. It shows that our approach is able to generate a hierarchy of different levels with valid and meaningful semantics.

Fig~\ref{fig:result_females} also shows another hierarchy for entities of people with a female gender extracted from the FB13k dataset~\cite{BordesUGWY13}. The results show a one level hierarchy where these entities are categorised into multiple groups. For example, the group of those having an education in the Barnard College is a subset of the group of people with a female gender. We have investigated on the Barnard college and found out that it is a private women's liberal arts college located in Manhattan, New York City. Founded in 1889 by Annie Nathan Meyer, who named it after Columbia University's 10th president, Frederick Barnard, it is one of the oldest women's colleges in the world. Similarly for other associated colleges in Fig.~\ref{fig:result_females}, these colleges are all women colleges.

We have also published further outcomes and hierarchy views on a publicly available website~\footnote{\url{https://samehkamaleldin.github.io/kg-hierarchies-gallery/}}.

\subsection{Implementation}
The experiments are implemented in Python3.5, and all the experiments were executed on a Linux machine with an Intel(R) Core(TM) i70.4790K CPU @ 4.00GHz and 32 GB RAM. We also use the D3 JavaScript library to visualise our hierarchy in a radial tree form.

\section{Related Work}
\label{sec:related}

Despite the widespread uses of knowledge graph in multiple domains, they suffer from missing information, especially type-based assertions of their entities. Multiple works were developed to tackle this problem including classical link prediction models~\cite{DBLP:journals/pieee/Nickel0TG16}, where multiple model use graph features~\cite{Mohamed2018KnowledgeBC} and embeddings~\cite{DBLP:journals/tkde/WangMWG17} to learn type links in the knowledge graphs. Also, type assertions of entities can be learnt using association rule mining models~\cite{Galrraga2013AMIEAR,Mohamed2017IdentifyingER} that identify type-based rules in the graph.

Further works have also focused on developing methods that exclusively predict entity types in knowledge graphs. These models have utilised different techniques including schema-based inference of type information~\cite{DBLP:conf/rweb/PolleresHDU13} and combination of ontologies and graph patterns~\cite{DBLP:conf/semweb/PaulheimB13}. Furthermore, other techniques have utilised machine learning models to infer types where they learn a feature representation of entities and their known types~\cite{DBLP:journals/ijswis/PaulheimB14,DBLP:conf/wims/MeloPV16}, and use these learnt features to infer new type link for other untyped entities in the knowledge graph.

\section{Conclusions and Future Work}
\label{sec:conc}
In this work, we have discussed the problem on knowledge graph entity classification, and we have shown that current state-of-the-art solutions are limited. We have also proposed a new approach for hierarchical grouping for knowledge graph entities which utilises an intuitive grouping of entities connected with the same predicate object combinations. We have shown that our approach can provide named hierarchical categorisation for the knowledge graph entities in a scalable parallel procedure. Our approach also operates on noisy data data by using flexible similarity measures. We have experimented our approach on standard knowledge graph benchmarking datasets and we have published the outcome hierarchies.

In future works we intend to add the new generated groups as triples to the graph along with their equivalence and dependence relationships and evaluate their effects on tasks like link prediction on knowledge graphs. We also intend to examine the possible use of the outcome hierarchies in mining association rules in knowledge graphs.

\section{Acknowledgements}
This work has been supported by Insight Centre for Data Analytics at National University of Ireland Galway, Ireland (supported by the Science Foundation Ireland grant 12/RC/2289).

\bibliographystyle{unsrt}
\bibliography{literature.bib}

\begin{thebibliography}{10}

\bibitem{DBLP:conf/www/MengCMSZ15}
Changping Meng, Reynold Cheng, Silviu Maniu, Pierre Senellart, and Wangda
  Zhang.
\newblock Discovering meta-paths in large heterogeneous information networks.
\newblock In {\em {WWW}}, pages 754--764. {ACM}, 2015.

\bibitem{DBLP:conf/semweb/KrompassBT15}
Denis Krompa{\ss}, Stephan Baier, and Volker Tresp.
\newblock Type-constrained representation learning in knowledge graphs.
\newblock In {\em International Semantic Web Conference {(1)}}, volume 9366 of
  {\em Lecture Notes in Computer Science}, pages 640--655. Springer, 2015.

\bibitem{DBLP:journals/kbs/ShiW16}
Baoxu Shi and Tim Weninger.
\newblock Discriminative predicate path mining for fact checking in knowledge
  graphs.
\newblock {\em Knowl.-Based Syst.}, 104:123--133, 2016.

\bibitem{DBLP:conf/ijcai/XieLS16}
Ruobing Xie, Zhiyuan Liu, and Maosong Sun.
\newblock Representation learning of knowledge graphs with hierarchical types.
\newblock In {\em {IJCAI}}, pages 2965--2971. {IJCAI/AAAI} Press, 2016.

\bibitem{DBLP:conf/wims/MeloPV16}
Andr{\'{e}} Melo, Heiko Paulheim, and Johanna V{\"{o}}lker.
\newblock Type prediction in {RDF} knowledge bases using hierarchical
  multilabel classification.
\newblock In {\em {WIMS}}, pages 14:1--14:10. {ACM}, 2016.

\bibitem{DBLP:journals/ijswis/PaulheimB14}
Heiko Paulheim and Christian Bizer.
\newblock Improving the quality of linked data using statistical distributions.
\newblock {\em Int. J. Semantic Web Inf. Syst.}, 10(2):63--86, 2014.

\bibitem{DBLP:conf/rweb/PolleresHDU13}
Axel Polleres, Aidan Hogan, Renaud Delbru, and J{\"{u}}rgen Umbrich.
\newblock {RDFS} and {OWL} reasoning for linked data.
\newblock In {\em Reasoning Web}, volume 8067 of {\em Lecture Notes in Computer
  Science}, pages 91--149. Springer, 2013.

\bibitem{DBLP:conf/semweb/PaulheimB13}
Heiko Paulheim and Christian Bizer.
\newblock Type inference on noisy {RDF} data.
\newblock In {\em International Semantic Web Conference {(1)}}, volume 8218 of
  {\em Lecture Notes in Computer Science}, pages 510--525. Springer, 2013.

\bibitem{Lu2010LinkPI}
Linyuan Lu and Tao Zhou.
\newblock Link prediction in complex networks: A survey.
\newblock {\em CoRR}, abs/1010.0725, 2010.

\bibitem{Miller:1995}
George~A. Miller.
\newblock {WordNet}: {A} lexical database for english.
\newblock {\em Commun. {ACM}}, 38(11):39--41, 1995.

\bibitem{DBLP:conf/nips/BordesUGWY13}
Antoine Bordes, Nicolas Usunier, Alberto Garc{\'{\i}}a{-}Dur{\'{a}}n, Jason
  Weston, and Oksana Yakhnenko.
\newblock Translating embeddings for modeling multi-relational data.
\newblock In {\em {NIPS}}, pages 2787--2795, 2013.

\bibitem{DBLP:conf/aaai/DettmersMS018}
Tim Dettmers, Pasquale Minervini, Pontus Stenetorp, and Sebastian Riedel.
\newblock Convolutional 2d knowledge graph embeddings.
\newblock In {\em {AAAI}}. {AAAI} Press, 2018.

\bibitem{DBLP:conf/sigmod/BollackerEPST08}
Kurt~D. Bollacker, Colin Evans, Praveen Paritosh, Tim Sturge, and Jamie Taylor.
\newblock Freebase: a collaboratively created graph database for structuring
  human knowledge.
\newblock In {\em {SIGMOD} Conference}, pages 1247--1250. {ACM}, 2008.

\bibitem{DBLP:conf/emnlp/ToutanovaCPPCG15}
Kristina Toutanova, Danqi Chen, Patrick Pantel, Hoifung Poon, Pallavi
  Choudhury, and Michael Gamon.
\newblock Representing text for joint embedding of text and knowledge bases.
\newblock In {\em {EMNLP}}, pages 1499--1509. The Association for Computational
  Linguistics, 2015.

\bibitem{DBLP:conf/cidr/MahdisoltaniBS15}
Farzaneh Mahdisoltani, Joanna Biega, and Fabian~M. Suchanek.
\newblock {YAGO3:} {A} knowledge base from multilingual wikipedias.
\newblock In {\em {CIDR}}. www.cidrdb.org, 2015.

\bibitem{DBLP:conf/aaai/Bouchard15}
Guillaume Bouchard, Sameer Singh, and Th{\'{e}}o Trouillon.
\newblock On approximate reasoning capabilities of low-rank vector spaces.
\newblock In {\em {AAAI} Spring Syposium on Knowledge Representation and
  Reasoning (KRR): Integrating Symbolic and Neural Approaches}. {AAAI} Press,
  2015.

\bibitem{DBLP:journals/cacm/MitchellCHTYBCM18}
Tom~M. Mitchell, William~W. Cohen, Estevam R.~Hruschka Jr., Partha~P. Talukdar,
  Bo~Yang, Justin Betteridge, Andrew Carlson, Bhavana~Dalvi Mishra, Matt
  Gardner, Bryan Kisiel, Jayant Krishnamurthy, Ni~Lao, Kathryn Mazaitis, Thahir
  Mohamed, Ndapandula Nakashole, Emmanouil~A. Platanios, Alan Ritter, Mehdi
  Samadi, Burr Settles, Richard~C. Wang, Derry Wijaya, Abhinav Gupta, Xinlei
  Chen, Abulhair Saparov, Malcolm Greaves, and Joel Welling.
\newblock Never-ending learning.
\newblock {\em Commun. {ACM}}, 61(5):103--115, 2018.

\bibitem{DBLP:conf/emnlp/GardnerM15}
Matt Gardner and Tom~M. Mitchell.
\newblock Efficient and expressive knowledge base completion using subgraph
  feature extraction.
\newblock In {\em Proceedings of the 2015 Conference on Empirical Methods in
  Natural Language Processing, {EMNLP} 2015, Lisbon, Portugal, September 17-21,
  2015}, pages 1488--1498, 2015.

\bibitem{BordesUGWY13}
Antoine Bordes, Nicolas Usunier, Alberto Garc{\'{\i}}a{-}Dur{\'{a}}n, Jason
  Weston, and Oksana Yakhnenko.
\newblock Translating embeddings for modeling multi-relational data.
\newblock In {\em Advances in Neural Information Processing Systems 26: 27th
  Annual Conference on Neural Information Processing Systems 2013. Proceedings
  of a meeting held December 5-8, 2013, Lake Tahoe, Nevada, United States.},
  pages 2787--2795, 2013.

\bibitem{DBLP:journals/pieee/Nickel0TG16}
Maximilian Nickel, Kevin Murphy, Volker Tresp, and Evgeniy Gabrilovich.
\newblock A review of relational machine learning for knowledge graphs.
\newblock {\em Proceedings of the {IEEE}}, 104(1):11--33, 2016.

\bibitem{Mohamed2018KnowledgeBC}
Sameh~K. Mohamed, V{\'i}t Nov{\'a}cek, and Pierre-Yves Vandenbussche.
\newblock Knowledge base completion using distinct subgraph paths.
\newblock In {\em SAC}, 2018.

\bibitem{DBLP:journals/tkde/WangMWG17}
Quan Wang, Zhendong Mao, Bin Wang, and Li~Guo.
\newblock Knowledge graph embedding: {A} survey of approaches and applications.
\newblock {\em {IEEE} Trans. Knowl. Data Eng.}, 29(12):2724--2743, 2017.

\bibitem{Galrraga2013AMIEAR}
Luis Gal{\'a}rraga, Christina Teflioudi, Katja Hose, and Fabian~M. Suchanek.
\newblock Amie: association rule mining under incomplete evidence in
  ontological knowledge bases.
\newblock In {\em WWW}, 2013.

\bibitem{Mohamed2017IdentifyingER}
Sameh~K. Mohamed, Emir Mu{\~n}oz, V{\'i}t Nov{\'a}cek, and Pierre-Yves
  Vandenbussche.
\newblock Identifying equivalent relation paths in knowledge graphs.
\newblock In {\em LDK}, 2017.

\end{thebibliography}

\end{document}